\documentclass[pre,12pt]{revtex4}

\usepackage{graphicx}
\usepackage{latexsym}
\usepackage[centertags]{amsmath}
\usepackage{amsfonts}
\usepackage{amssymb}
\usepackage{amsthm}
\usepackage{newlfont}
\usepackage{color}

\begin{document}
 
\title{Scaling properties of field-induced superdiffusion \\ in Continous Time Random Walks}

\author{R. Burioni} 
\affiliation{Dipartimento di Fisica e Scienza
  della Terra, Universit\`a degli Studi di Parma and \\ INFN, Gruppo
  Collegato di Parma, viale G.P.Usberti 7/A, 43126 Parma, Italy}
\affiliation{Kavli Institute for Theoretical Physics, Beijing 100190, China}

\author{G. Gradenigo} \affiliation{Institut de Physique Th\'eorique, CEA Saclay,
  F-91191 Gif-sur-Yvette Cedex, France} 
\affiliation{Laboratoire de
  Physique Th\'eorique et Mod\`eles Statistiques, CNRS et Universit\'e
  Paris-Sud 11, 91405 Orsay Cedex, France}

\author{A. Sarracino}
\affiliation{ISC-CNR and Dipartimento di Fisica, Universit\`a
  Sapienza, p.le A. Moro 2, 00185 Roma, Italy} 
\affiliation{Kavli Institute for Theoretical Physics, Beijing 100190, China}
\affiliation{Laboratoire de Physique Th\'{e}orique de la Mati\`{e}re Condens\'{e}e, CNRS, Universit\'e Paris 6, 4 Place Jussieu, 75255 Paris Cedex}

\author{A. Vezzani}
\affiliation{Centro S3, CNR-Nanoscienze, Via Campi 213A,
  41125 Modena, Italy and \\ Dipartimento di Fisica e Scienza della
  Terra, Universit\`a degli Studi di Parma, viale G.P.Usberti 7/A,
  43126 Parma, Italy} 

\author{A. Vulpiani} \affiliation{Dipartimento
  di Fisica, Universit\`a Sapienza and ISC-CNR, p.le A. Moro 2, 00185
  Roma, Italy}
\affiliation{Kavli Institute for Theoretical Physics China, CAS, Beijing 100190, China}

\begin{abstract}
{\bf Abstract:} We consider a broad class of
Continuous Time Random Walks with large fluctuations effects in
space and time distributions: a random walk with trapping, describing
subdiffusion in disordered and glassy materials, and a L\'evy walk
process, often used to model superdiffusive effects in inhomogeneous materials. 
 We derive the scaling form of the probability distributions and
the asymptotic properties of all its moments  in the presence of  a field by two powerful techniques, based on matching conditions and on the estimate of the contribution of rare events to power-law tails in  a field. 
\end{abstract}

\maketitle

\section{Introduction}

Physical processes occurring in a wide class of phenomena are strongly affected by large fluctuations and rare events. These are often crucial to determine the behavior of a system and their contribution and their statistics strongly influence the observed physical properties. Large fluctuation effects have been observed in laser cooling~\cite{bardou}, in liposome
diffusion~\cite{wang}, in heteropolymers dynamics~\cite{tang}, in 
systems with chaos~\cite{sanchez} and in non-equilibrium
relaxation~\cite{bouchaud,ribiere}. In these systems, the typical heterogeneity
of spatial structures or temporal patterns can lead to
\emph{anomalous} transport: the mean square displacement (MSD) of a
particle is not linear in time, so that $\langle x^{2}(t)\rangle\sim
t^{2/\gamma}$, with $\gamma \neq 2$, and one observes
 subdiffusion, $\gamma>2$, or superdiffusion, $\gamma<2$.  However, if
the fluctuations in the microscopic dynamics are sufficiently small,
standard diffusion can survive also in the presence of heterogeneity. 

In these subtle situations, one can still be able to detect the
heterogeneous nature of the underlying microscopic dynamic by means of
an applied field.  Indeed, recently, we have analyzed a broad class of
Continuous Time Random Walks (CTRW) with large fluctuations effects in
space and time distributions: a random walk with trapping, typical of
subdiffusion in disordered and glassy materials, and a L\'evy walk
process, often used to model superdiffusion in inhomogeneous media
\cite{noi}. It has been shown that in presence of an external field an
anomalous growth of fluctuations can be observed even when the
unperturbed behavior is Gaussian \cite{beni1,beni3} and the Einstein
relation \cite{barkai,jespersen,villamaina,bettolo,beni2} holds.
Recent results of numerical simulations of complex
liquids~\cite{winter,heuer}, have shown that superdiffusion can be
induced by an external field in trap-like disordered systems.
However, as explained in~\cite{benill}, such a behavior is more
general, and can be observed in driven crowded systems, far from the
glass transition, due to confinement effects.

In our previous work, we have solved the master equations of the two
processes and we have determined the probability distribution $P_\epsilon(x,t)$ for the
particle to be at point $x$ at time $t$ in the
presence of an external field $\epsilon$.  Our results shows that when
transport is anomalous, the drift has a strong influence on
distributions, as expected. More surprisingly, in the regimes where
the form of the unperturbed distribution is Gaussian, the field can
induce a non-Gaussian shape of $P_\epsilon(x,t)$. In particular, in
the CTRW with trapping, in a region of the parameters of the highly
fluctuating waiting times distribution, the drift induces a superdiffusive
spreading of the perturbed non-Gaussian $P_\epsilon(x,t)$. Even in this subtle case,
the measure of the response to a field is therefore able to detect the anomalous
dynamics present in the waiting times and step lengths distribution.
Moreover, a small field can induce a transition from standard Gaussian
diffusion to a \emph{strong} anomalous one, i.e the scaling of moments
of different order is not described by a unique exponent: $\langle |x(t)|^q \rangle 
\sim t^{\nu(q)}$ with $\nu(q)\neq const \times q$ \cite{castiglione} .

The key ingredient to obtain our previous results relies on the scaling properties of
the probability distribution $P_\epsilon(x,t)$, and on the subtle effects of the different length scales present in the process in the presence of a field. Recently, different heuristic yet powerful techniques have been developed and applied to estimate the scaling  form of probability distribution functions: a matching condition arising from cut off effects ~\cite{ACMV00}, and the "single long jump hypothesis", 
that is able to extracts the largest contribution of rare events to the tails of the probability  distribution ~\cite{Levyrandom}.  In this paper, we apply these two interesting techniques to reproduce the scaling regimes observed in the two CTRW in their normal, anomalous and strong anomalous regimes.

Our models contain a scale invariant distribution of trapping times and a scale invariant
distributions of displacements. We consider a CTRW  where a Brownian particle is trapped for a time interval distributed according to  the
distribution ~\cite{bouchaud}:
\begin{equation}
  p_\alpha(\tau)\sim\tau^{-(\alpha+1)},
  \label{levy}
\end{equation}
where $\alpha>0$ is the exponent characterizing the slow decay ~\cite{angelani}.  

Then, we consider the case of fat tails in displacements distribution, that typically
arises in L\'evy-like motion in heterogeneous
 materials~\cite{levitz,havlin,Levy2d,Levy2drandom,beenakker2d,benichou}, or in turbulent flow~\cite{shles1985}
and granular materials ~\cite{lechenault}. Here, transport is realized through
increments of size $l$ with distribution $p_\alpha(l)\sim l^{-(1+\alpha)}$~\cite{barthelemy,klages} that can give rise to superdiffusive dynamics.

\section{Continuous Time Random Walks and the scaling hypotheses}

Let us now discuss in details the two models and the scaling forms of the probability distributions. The CTRW with trapping describes a particle moving with probability $1/2$ from $x$ to $x \pm \delta_0$, with $\delta_0$ constant, or extracted from 
a symmetric distribution with finite variance. Between successive
steps, the particle is trapped for a time $\tau$ extracted from the
distribution~(\ref{levy}).  In the L\'evy walk model, during the time
interval $\tau$, always extracted from the same distribution~(\ref{levy}), the
particle moves at constant velocity $v$. The velocity is here chosen from a symmetric
distribution with finite variance $p(v)$, so that the particle performs displacements
$\delta=\tau v$. We consider a Gaussian distribution $p(v)$ for the velocity. In
both models, we introduce a lower cutoff $\tau_0=1$ in the
distribution~(\ref{levy}) so that, taking into account the
normalization, we have
\begin{equation}
p_\alpha(\tau)=\alpha \Theta(\tau-1) \tau^{-(1+\alpha)},
\label{pditau}
\end{equation}
where $\Theta(x)$ is the Heaviside step function.
The value of $\tau_0$ does not change the behavior of the models apart from a 
suitable rescaling of the constants.

In the CTRW with trapping, the external field unbalances the jump probabilities, i.e. setting to $(1\pm\epsilon)/2$
the probability of jumping to the right or to the left,
respectively. For the L\'evy walk, the system is driven out of equilibrium by applying an external field accelerating the particle during the
walk, so that the distance travelled in the time interval $\tau$ is
$\delta= \pm v_0 \tau + \epsilon \tau^2$. In the following, we will
consider a positive bias $\epsilon>0$ \cite{sokolov,gradenigo} .

When diffusion is normal and the probability distribution is Gaussian, the simplest scaling form of
the probability density function $P_\epsilon(x,t)$, for $\epsilon\ne 0$ can be written
as
\begin{equation} 
P_\epsilon(x,t) \sim t^{-1/z} F[(x-v_\epsilon t)/t^{1/z}].
\label{scaling}
\end{equation}
When $\epsilon=0$, the left-right symmetry implies that $v_\epsilon=0$
and $F(y)$ is an even function.  

Our results shows that the scaling form of the distribution ~(\ref{scaling}) becomes more complex in the two models we describe here, because of rare events. For the L\'evy walks, these
rare events are large displacements in the direction of the field, while  in the CTRW with trapping, they correspond to large trapping times of particles trapped  at  small values of $x$ \cite{noi}.  Due to these effects, the scaling form of
Eq.~(\ref{scaling}) has to be modified, as we must take into account
the cut-off in the largest distance from the peak of the
distribution that the particle can reach.  In the
L\'evy walk at time $t$, there can be no displacement exceeding $v_0 t
+ \epsilon t^2$, so that for large times we have:
\begin{equation} 
P_\epsilon(x,t) \sim t^{-1/z} F[(x-v_\epsilon
  t)/t^{1/z}]~\Theta\left(\epsilon t^2-x\right).
\label{scaling2}
\end{equation}
In the CTRW with trapping, the factor $\Theta(\epsilon t^2-x)$ in
Eq.~(\ref{scaling2}) has to be replaced by $\Theta(x)$, which cuts off
the power-law tail due to particles with large persistence time at the
origin.  In Eq.~(\ref{scaling2}) we can define three characteristic
lengths: the length-scale of the peak displacement, $l_T(t) \sim t $;
the length-scale for the collapse of the central part of the probability distribution function, $l(t)\sim t^{1/z}$;
the length-scale of the largest displacement from the peak of the
distribution, $l_{e}(t) \sim t^2$. The presence of these additional length scales
strongly modifies the behavior of the mean square displacement and of all the higher moments.
Hereafter, we will call $x_p(t)\sim \ell_T(t)$ the position
of the peak of the distribution and $\xi= x- x_p(t)$ the distance from 
this peak.

Let us now see how the scaling form of the distribution and the asymptotic behavior of its moments can be obtained by two powerful techniques: a matching condition in presence of cut offs and the "single long jump hypothesis".

\section{Rare events and scaling in CTRW with trapping}

\subsection{Scaling of moments from the matching argument}

Let us first consider the CTRW with trapping. A simple scaling argument~\cite{ACMV00} with a matching condition can be used to predict the asymptotic
behavior of the second moments of the displacements and to recover the regimes of anomalous and strong anomalous diffusion \cite{noi}.

Assuming $x=0$ as the initial condition for each trajectory, the mean
position at time $t$, after $N(t)$ jumps have
occurred, is:
\begin{equation}
\langle x \rangle_\epsilon = \left\langle \left[ \sum_{i=1}^{N(t)} x_i \right] \right\rangle_\epsilon 
\label{pos0}
\end{equation}
while the mean square displacement is given by:
\begin{equation}
\langle x^2(t) \rangle_\epsilon = \left\langle \left[ \sum_{i=1}^{N(t)} x_i \right]^2 \right\rangle_\epsilon 
\label{MSD0}
\end{equation}
Here, $x_i$ denotes the displacement of the particle at jump $i$. At
zero field, the displacement takes values $\pm 1$ with probability $1/2$, so that
$\langle x_i\rangle_0=0$. When the field $\epsilon$ is turned on, 
the jumping probabilities are unbalanced 
so that $\langle x_i\rangle_\epsilon=\epsilon$. Analogously,
the fluctuations are $\Delta(x)^2=\langle x^2\rangle_\epsilon-\langle x\rangle^2_\epsilon=1+\epsilon^2$

In order to evaluate the sum in Eqs.~(\ref{pos0},\ref{MSD0}) 
we must take into account the fluctuations of the step lengths and
also the fluctuations of the number of steps, and therefore we write:
\begin{equation} 
\langle x(t) \rangle_\epsilon = \left\langle  \sum_{i=1}^{N(t)} x_i  \right\rangle_\epsilon = 
\overline{N(t)} \langle x_i \rangle_\epsilon,
\label{pos0trap}
\end{equation} 
\begin{equation} 
\langle x^2(t) \rangle_\epsilon = \left\langle  \sum_{i=1}^{N(t)} x_i  \sum_{j=1}^{N(t)} x_j \right\rangle_\epsilon = 
\overline{N(t)} \left( \langle x_i^2 \rangle_\epsilon - \langle x_i \rangle_\epsilon^2 \right)+ \overline{N(t)^2} \langle x_i \rangle^2_\epsilon 
\label{moment0trap}
\end{equation} 
and
\begin{equation} 
\langle x^2(t) \rangle_\epsilon - \langle x(t) \rangle_\epsilon^2 = 
\overline{N(t)} \left( \langle x_i^2 \rangle_\epsilon - \langle x_i \rangle_\epsilon^2 \right)+ \left(\overline{N(t)^2} - \overline{N(t)}^2 \right)\langle x_i \rangle^2_\epsilon=\overline{N(t)} \Delta_\epsilon(x_i)^2 + \Delta(N)^2\langle x_i \rangle^2_\epsilon
\label{flut0trap}
\end{equation} 
where $\overline{N(t)}$ is the average number of steps in a time 
$t$ and $\Delta(N)$ represents its fluctuations.

Let us first focus on the case  $\alpha>2$, when the waiting 
time distribution $p_\alpha(\tau)$ displays a 
finite average $\langle \tau \rangle$ and a finite fluctuation 
$\Delta(\tau)^2 =\langle \tau^2 \rangle-\langle \tau \rangle^2$.
In this case, $N(t)$ is distributed according to an (inverse)-Gaussian with
mean $\overline{N(t)}=t/\langle \tau \rangle$ and variance
$\Delta(N)^2=t \Delta(\tau)^2/\langle \tau \rangle^3$. 
Plugging the above results in the formulas 
(\ref{moment0trap},\ref{flut0trap}),
we obtain the classical result of diffusion with drift:
$\langle x(t) \rangle = t \epsilon/\langle \tau \rangle$ and 
$\langle x^2(t) \rangle_\epsilon - \langle x(t) \rangle_\epsilon^2 \sim t$,
where the proportionality constant depends both on the mean and on the variance of the step length and of the waiting time distributions.

In the case of $\alpha <2$, the behaviour of 
 $\overline{N(t)}$ and  $\Delta(N)$ can be recovered 
in a simple way introducing a cutoff
$t_c$ for $p_\alpha(\tau)$ ~\cite{ACMV00}:
\begin{equation}\label{cutoff}
p_\alpha(\tau) \sim \left\{ \begin{array}{ll}
p_\alpha(\tau) & \textrm{if $\tau<t_c$}\\
0 & \textrm{if $\tau>t_c$.}\\ 
\end{array} \right.
\end{equation}
so that the variance and the fluctuations remain finite. Again, we have
$\overline{N(t)}=t/\langle \tau \rangle_c$ and 
$\Delta(N)=t \Delta_c(\tau)^2/\langle \tau \rangle_c^3$
where, with the index $c$, we denoted quantities evaluated taking into 
account of the cutoff at $t_c$.
Then, the correct asymptotic behavior can be obtained by letting the cutoff
$t_c \rightarrow t$, i.e. the maximum waiting time allowed in a process of 
duration $t$ ~\cite{ACMV00}. 

With this procedure, for $1<\alpha<2$ we obtain
$\langle \tau \rangle_c =\langle \tau \rangle=const$ and 
$\Delta_c(\tau)^2\sim t^{2-\alpha}$. Using
this results in Eq. (\ref{moment0trap},\ref{flut0trap}) we have,
for $\epsilon=0$, $\langle x(t) \rangle=0$ and
$\langle x(t)^2 \rangle \sim t$, that is the typical behavior of normal diffusion. 
However, for $\epsilon\not=0$ a strong anomalous behavior arises, 
as already discussed in \cite{noi}:
\begin{eqnarray} 
\langle x (t) \rangle_\epsilon &=&  \epsilon t/\langle \tau \rangle 
\nonumber \\ 
\langle x^2 (t) \rangle_\epsilon & \sim &   t^{3-\alpha} \langle x_i\rangle_\epsilon^2.
\label{moment1trap}
\end{eqnarray}

Finally, we focus on the case $\alpha<1$. Now for $t_c\rightarrow t$ 
we get $\langle \tau \rangle_c \sim  t^{1-\alpha}$ and 
$\Delta_c(\tau)\sim t^{2-\alpha}$. 
From Eq. (\ref{moment0trap},\ref{flut0trap}), we obtain for $\epsilon =0$:
\begin{eqnarray} 
\langle x (t) \rangle_0 &=&  0 \nonumber \\ 
\langle x^2 (t) \rangle_0 & \sim &   t^{\alpha} \Delta_0(x_i)^2.
\label{moment2trap}
\end{eqnarray}
while when $\epsilon \not=0$ we have:
\begin{eqnarray} 
\langle x (t) \rangle_\epsilon &\sim&   t^{\alpha}  \nonumber \\ 
\langle x^2 (t) \rangle_\epsilon-\langle x (t) \rangle_\epsilon^2 & \sim &   t^{2\alpha}.
\label{moment3trap}
\end{eqnarray}
These equations implies that in this case diffusion is anomalous but not
strongly anomalous \cite{noi}. Indeed we have  
$\langle [\delta x(t)]^2 \rangle_\epsilon 
\sim \langle x(t) \rangle_\epsilon^2 \sim \ell(t)^2$. 
Moreover we recover the results of~\cite{bouchaud}, which
predicts a superdiffusive field-induced dynamics in the case $1/2 <
\alpha < 1$.

\subsection{Numerical results for the probability distribution: $\alpha < 1/2$ } 

\begin{figure}
  \includegraphics[width=.45\columnwidth,clip=true]{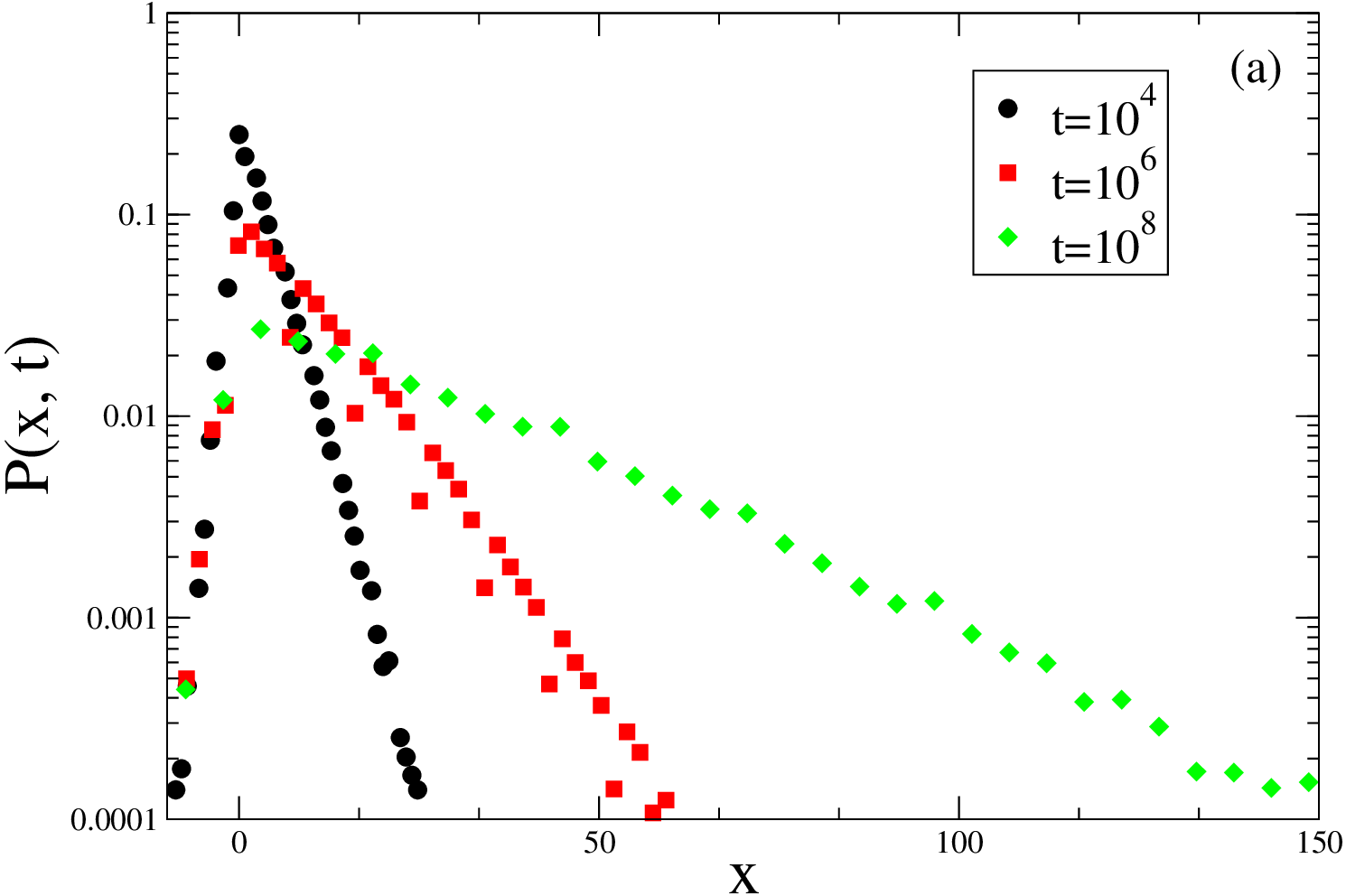}
  \includegraphics[width=.45\columnwidth,clip=true]{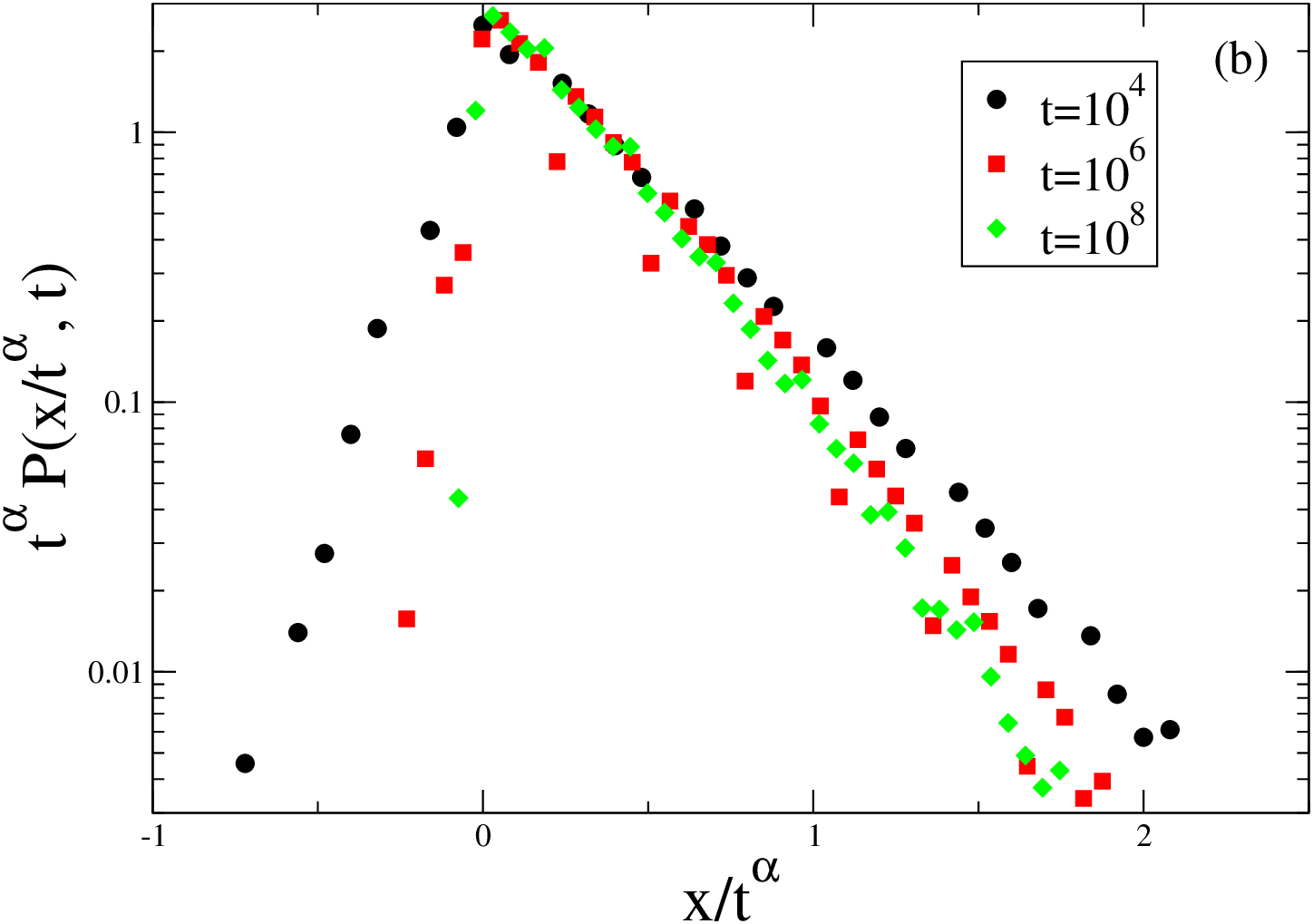}
  \caption{Probability distribution of displacements $P_\epsilon(x,t)$
    for an asymmetric CTRW (at each step the probability to move left
    or right is respectively $p_l = (1-\epsilon)/2$ and $p_r =
    (1+\epsilon)/2$, with $\epsilon=0.3$) with exponent {\bf $\alpha =
      1/4 $} of waiting time distribution $p(\tau) =
    \tau^{-(1+\alpha)}$. {\bf Left:} $P_\epsilon(x,t)$ at
    $t=10^4,10^6,10^8$ in semilog scale; {\bf Right:}
    $P_\epsilon(x,t)$ at $t=10^4,10^6,10^8$ in semilog scale with
    data collapsed according to the scaling form $P_\epsilon(x,t) =
    t^{-\alpha} F_\epsilon(x/t^\alpha)$.}
  \label{fig:trap-alpha0.25}
\end{figure}

We now check our scaling picture for the probability distribution $P_\epsilon(x,t)$ in the different regimes of the CTRW with trapping.
As already shown, in the regime $\alpha < 1/2$, the scaling length of the process is $t^\alpha$
and the effect of the external field is not sufficient to
induce a ``superdiffusive'' spreading of the probability distribution
of displacements: as we can see from 
Eq.~(\ref{moment3trap}) both the drift and the mean square
displacement around the drift for $\alpha <1/2$ are subdiffusive. As
reported in~\cite{villamaina}, for the case $\alpha = 1/2 $ this
corresponds to a distribution $P_\epsilon(x,t)$ which,
despite the positive field, is not shifting to the right. 
In this case $P_\epsilon(x,t)$ develops a long positive tail, which slowly
increases with time, as $\langle x (t) \rangle_\epsilon
\sim t^{\alpha}$ (see Fig.\ref{fig:trap-alpha0.25}). From Fig. (\ref{fig:trap-alpha0.25}),
one sees that the peak of the distribution is centered in $x_p(t)=0$ 
and the curves show a very good scaling.

\begin{figure}
  \includegraphics[width=.45\columnwidth,clip=true]{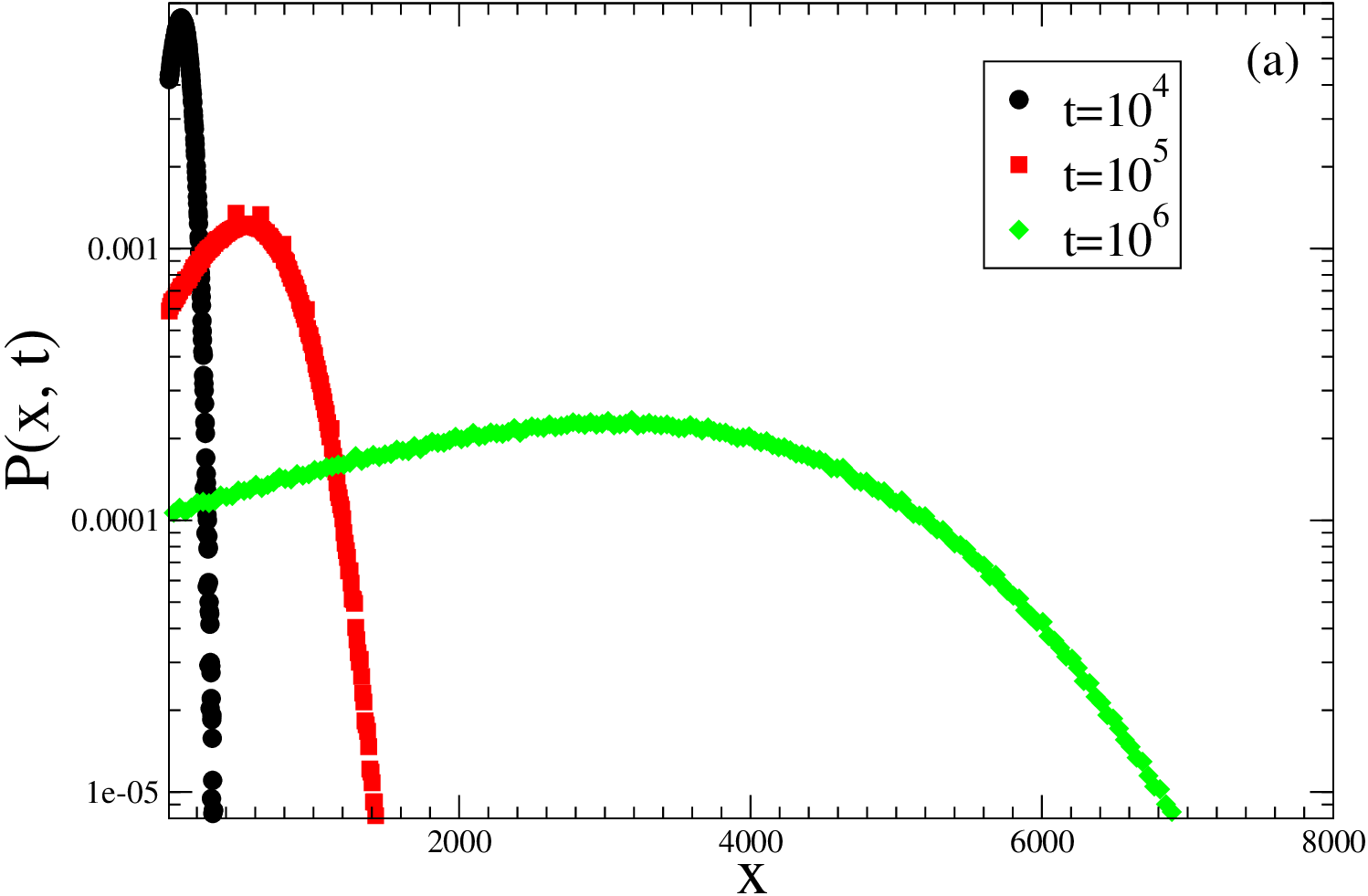}
  \includegraphics[width=.45\columnwidth,clip=true]{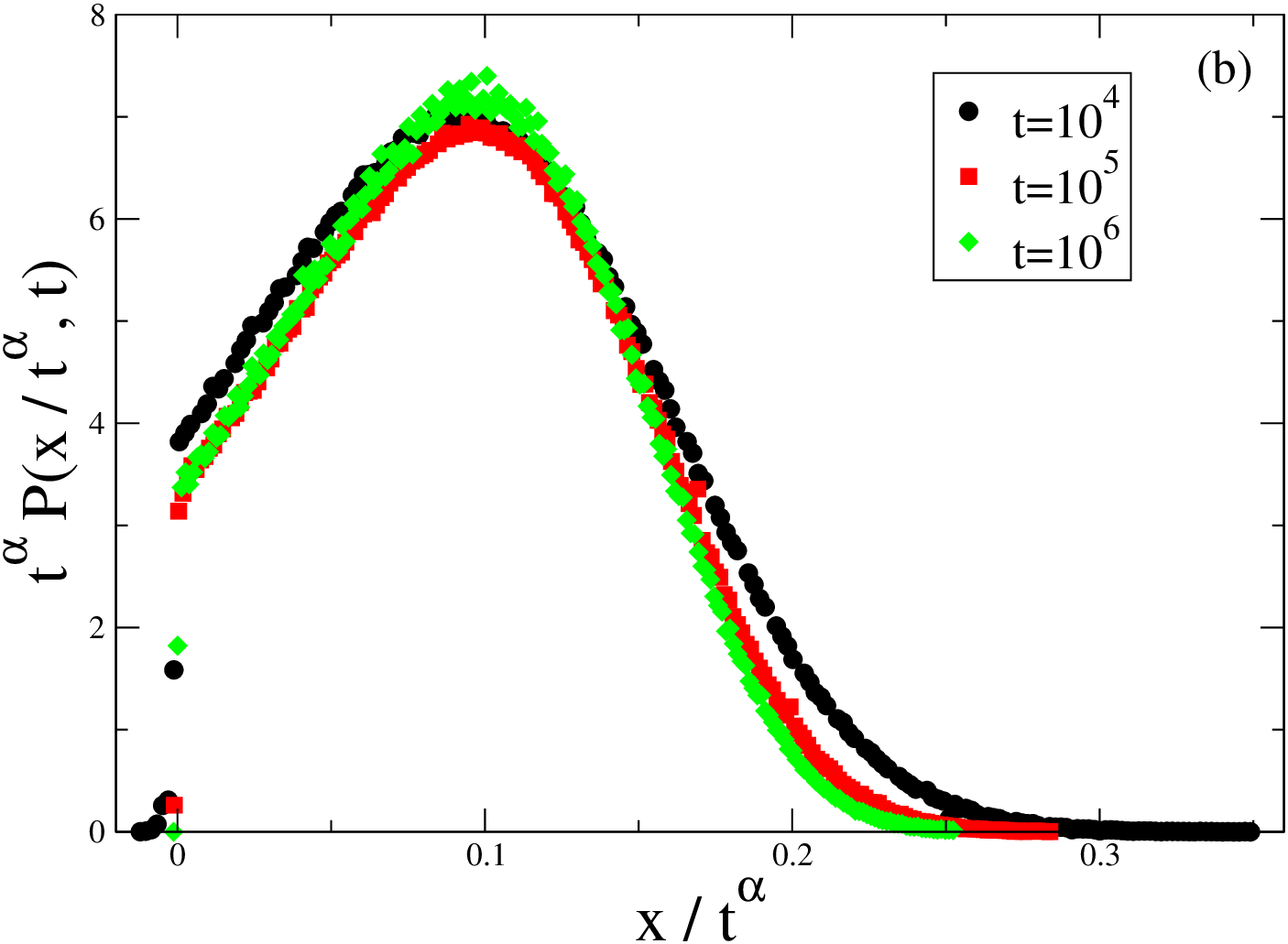}
  \caption{Probability distribution of displacements $P_\epsilon(x,t)$
    for an asymmetric CTRW (at each step the probability to move left
    or right is respectively $p_l = (1-\epsilon)/2$ and $p_{r} =
    (1+\epsilon)/2$, with $\epsilon=0.3$) with exponent {\bf $\alpha =
      3/4$} of waiting time distribution $p(\tau) =
    \tau^{-(1+\alpha)}$. {\bf Left:} $P_\epsilon(x,t)$ at
    $t=10^4,10^5,10^6$ in semilog scale; {\bf Right:}
    $P_\epsilon(x,t)$ at $t=10^4,10^5,10^6$ in linear scale with data
    collapsed according to the scaling form $P_\epsilon(x,t) =
    t^{-\alpha} F_\epsilon(x/t^\alpha)$.}
  \label{fig:trap-alpha0.75}
\end{figure}

\subsection{Numerical results for the probability distribution: $1/2 < \alpha < 1$ }

The scaling picture for $1/2 < \alpha < 1$ is analogous to the case
with $\alpha < 1/2$, and the scaling length of the process is $\ell(t) \sim
t^{\alpha}$. In this situation we observe the onset of a field-induced
superdiffusive spreading of $P_\epsilon(x,t)$ in a regime where in absence of
an external field the behavior is still subdiffusive, $\langle x(t)^2 \rangle_0 \sim t^\alpha$. 
The main difference with respect to the previous case is that now the peak of the distribution
moves as $x_p(t)\sim t^{\alpha}\sim\ell(t)$ and the scaling form reads:

\begin{equation}
P(x,t)\sim \frac{1}{t^\alpha} F\left( \frac{\xi}{t^\alpha} \right) \sim \frac{1}{t^\alpha} F\left( \frac{|x- v_{sub} t^\alpha|}{t^\alpha} \right) \sim
\frac{1}{t^\alpha} G\left( \frac{x}{t^\alpha} \right) 
\label{scaling1btrap}
\end{equation}
where $\alpha>1/2$. Fig.~\ref{fig:trap-alpha0.75} shows that the scaling
assumption in Eq.~(\ref{scaling1btrap}) is well verified. 
Interestingly, as clearly seen from Fig.s~\ref{fig:trap-alpha0.25} and 
\ref{fig:trap-alpha0.75},
while before the rescaling the behavior of the probability distribution 
at $1/2 < \alpha <1$ looks very different
from that observed in the case $\alpha < 1/2$, after rescaling the two curves
display the same behavior and the only difference is the position of the 
peak at $x=0$ and at $x>0$.  
Also in this regime,  we observe a \emph{weak} anomalous
diffusion, because there is a single scaling length in the
problem, $\ell(t)=t^\alpha$, and all the higher order cumulants can be
written as functions of this scaling length: $\langle \left[ \delta
  x(t)\right]^n \rangle_\epsilon \sim \ell^n(t)$.

\subsection{Numerical results for the probability distribution: $1 < \alpha < 2$}

For $1 < \alpha < 2$ the drift of the  distribution $P_\epsilon(x,t)$
becomes linear, $x_p(t) \sim \langle x(t)\rangle_\epsilon \sim t$, 
and now the probability of finding a
particle at a distance $\xi=x-x_p(t)$ 
from the peak of the distribution at time $t$ 
can be evaluated with a simple scaling argument.
In particular the probability of remaining at a distance $\xi$ from
$x_p(t)$ is due to the probability for a walker to
experience a stop of duration $\tau\sim \xi$. 
For each scattering event the probability of waiting a time
$\tau\sim \xi$ is then
\begin{equation}   
p_\alpha(\tau) d\tau = p_\alpha(\tau(\xi)) \frac{d \tau}{d\xi} d \xi \sim 
\frac{1}{\xi^{1+\alpha}}d \xi, 
\end{equation}
Since the number of steps $N(t)$ in a time $t$ is also proportional to $t$,
the tail of the distribution at large $\xi$ can be estimated as: 
\begin{equation} 
P_\epsilon(\xi,t) \sim \frac{N(t)}{\xi^{1+\alpha}} \sim  
\frac{t}{\xi^{1+\alpha}},
\label{scaling2trap}
\end{equation}
which corresponds to a scaling form:  
\begin{equation} 
P_\epsilon(\xi,t) =
\frac{1}{t^{1/\alpha}} F\left( \frac{\xi}{t^{1/\alpha}} \right)=
\frac{1}{t^{1/\alpha}} F\left( \frac{|x- v t|}{t^{1/\alpha}} \right). 
\label{scaling2trapB}
\end{equation} 
Introducing the cut-off in zero to calculate the higher 
order cumulants of the $P_\epsilon(x,t)$, we obtain 
\begin{equation}
\langle \left[ \delta x(t)\right]^n \rangle_\epsilon = \frac{1}{t^{1/\alpha}}
\int_0^{t} d\xi ~\xi^n ~F\left( \frac{\xi}{t^{1/\alpha}} \right) \sim t \int_0^t d\xi ~ \xi^{n-1-\alpha} \sim t^{n+1-\alpha}.  
\label{scaling3trap}
\end{equation}
In the case of $1 < \alpha < 2$ the effect of a fat tail connecting
the center of the distribution to the origin is to produce a
\emph{strong} anomalous diffusion, because the scaling length of the
distribution is $\ell(t) \sim t^{\alpha/2}$ and from
Eq.~(\ref{scaling3trap}) we see that $\langle \left[ \delta x(t)\right]^n
\rangle_\epsilon \neq \ell^n(t)$.

\begin{figure}[!t]
  \includegraphics[width=.7\columnwidth,clip=true]{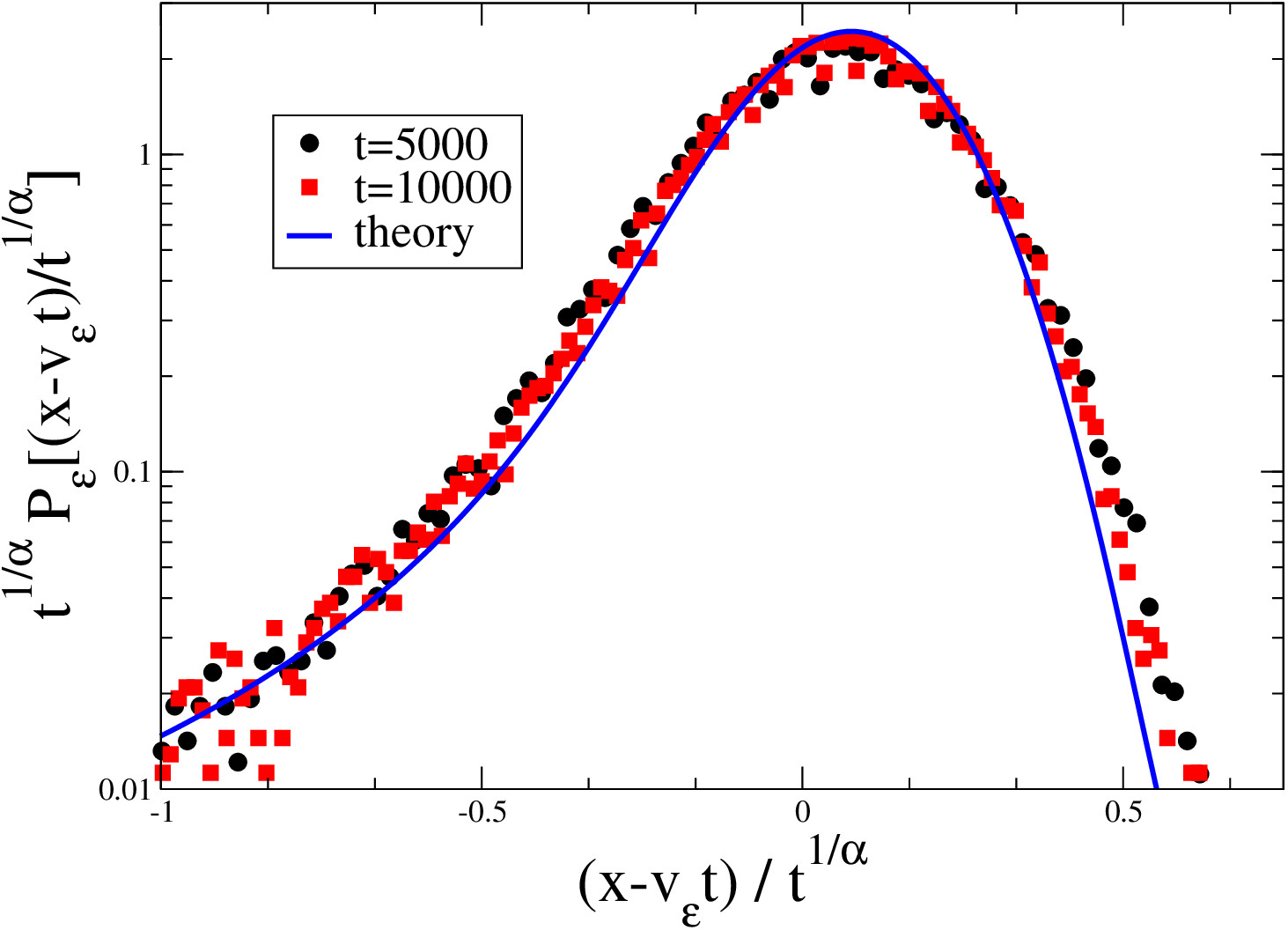}
  \caption{
    Probability distributions $P_\epsilon(x,t)$ of the Continuous Time
    Random Walk with trapping, calculated at different times in numerical simulations,
    are collapsed here according to the scaling form in Eq.~(\ref{scaling2trapB}),
    with exponent $\alpha=3/2$ and $v_\epsilon=0.1$. Symbols correspond to different times,
    as shown in the legend. The continuous line represents the analytical result reported in~\cite{noi}.}
  \label{fig:scaling-sub-alpha1.5}
\end{figure}

Detailed analytical computations are reported in~\cite{noi} and are in
very good agreement with the results of numerical simulations, as
shown in Fig.~\ref{fig:scaling-sub-alpha1.5} for $\alpha=1.5$.

\section{Rare events and scaling in the L\'evy Walk} 

\subsection{Scaling from the matching argument}

In the  L\'evy walk model  the probe particle
performs free flights with velocities $v_i$ extracted from a Gaussian
distribution, and of duration $\tau_i$, with a total displacement which
is, in absence of the field, $x(t) = \sum_i^{N(t)} \tau_i v_i$, where
$N(t)$ is the number of steps up to the elapsed time $t$. In this model
the external field $\epsilon$ acts as a constant acceleration, so that
the total displacement in a trajectory becomes $x(t) = \sum_i^{N(t)}
\tau_i v_i + \epsilon \tau_i^2 $. 
The mean value and the second moment of $x(t)$ can be evaluated analogously 
to the CTRW with trapping as:
\begin{equation} 
\langle x(t) \rangle_\epsilon = \left\langle  \sum_{i=1}^{N(t)} x_i  \right\rangle_\epsilon = 
\overline{N(t)} \langle \tau_i^2 \rangle_\epsilon ,
\label{pos0Levy}
\end{equation} 
and
\begin{equation} 
\langle x^2(t) \rangle_\epsilon - \langle x(t) \rangle_\epsilon^2 = 
\overline{N(t)} \left( \langle v_i^2 \rangle + \epsilon^2 (\langle \tau_i^4 \rangle_c - \langle \tau_i^2 \rangle_c^2) \right)+ \left(\overline{N(t)^2} + \Delta(N)^2 \right)\epsilon^2 \langle \tau_i \rangle^2_c, 
\label{flut0Levy}
\end{equation} 
where we used the fact that $\langle v_i \rangle=0$. Moreover,
in order to avoid divergences, we introduce the same
cutoff $t_c$ in the evaluation of  the average of the flights time 
and of its moments, using for this purpose again the 
modified distribution (\ref{cutoff}). 
Also in this case the average number of steps 
in a time $t$ and its variance can be evaluated as 
$N(t)=t/\langle \tau \rangle_c$ and 
$\Delta(N)^2=t \Delta_c(\tau)^2/\langle \tau \rangle_c^3$. Finally, the 
divergences can be taken safely into account letting the cutoff
$t_c \rightarrow t$ i.e. the maximum time of flight allowed in a time $t$ ~\cite{ACMV00} .

In the case $\alpha>4$ all the expectation values in Eqs. (\ref{pos0Levy},\ref{flut0Levy}) are finite and one recovers the standard formulas for diffusion 
with drift i.e. $\langle x(t) \rangle_\epsilon \sim t$ and 
$\langle x^2(t) \rangle_\epsilon - \langle x(t) \rangle_\epsilon^2 \sim t$. 
For $2<\alpha<4$ we get a standard behavior for the drift 
$\langle x(t) \rangle_\epsilon \sim t \epsilon$. However, in the evaluation 
of the variance 
we have to take into account that $\langle \tau_i^4 \rangle$ is diverging and,
setting the cut of at $t_c=t$, we have 
$\langle x^2(t) \rangle_\epsilon - \langle x(t) \rangle_\epsilon^2 
\sim t^{5-\alpha}\epsilon^2$. 
Therefore, the system features a strong anomalous behavior in a regime where 
in absence of forcing a normal diffusion is observed. In the regime 
$1<\alpha<2$, the variance has the same behavior but  an anomalous drift 
$\langle x(t) \rangle_\epsilon \sim t^{3-\alpha}$ is present, 
since $\langle \tau_i\rangle$ diverges as $t^{2-\alpha}$.
Finally for $\alpha<1$ the motion is accelerated $\langle x(t) \rangle_\epsilon \sim t^{2}$ and $\langle x^2(t) \rangle_\epsilon - \langle x(t) \rangle_\epsilon^2 \sim t^{4}$.

\subsection{Scaling from the single long jump hypothesis}

The matching argument allows us to evaluate the behavior of the first 
and second moment of the distribution  $P_\epsilon(x,t)$; 
we now introduce another simple
argument for estimating the scaling properties of $P_\epsilon(x,t)$
assuming that 
the behaviour of higher order moments/cumulants  is
due to a fat (right) tail and that the behaviour of the tail is
generated by isolated rare events ~\cite{Levyrandom}. 
This means that the probability
of finding a particle at a very large distance $\xi$
from $x_p(t)$, the peak of the distribution, 
 is given by the probability that this
particle has reached $\xi$ thanks to a single long
jump \cite{Levyrandom}. Therefore, due to the constant acceleration of 
the particle, the
large distribution of displacements in single jumps is related to the
large distribution of flight times through the change of variables
$\xi = v \tau + \epsilon/2 \tau^2$, which for large time can be
approximated as $\xi \sim \tau^2 $. From this we can obtain the
distribution of displacements in a single jump in presence of the
field:
\begin{equation}
p(\tau) d\tau= p(\tau(\xi)) \frac{d \tau}{d\xi} d \xi \sim 
\frac{1}{\xi^{1+\alpha/2}}d \xi. 
\end{equation} 
From this relation, we can predict that the general scaling form 
of the fat tail of $P_\epsilon(x,t)$
will be related to trajectories dominated by a single long jump $\xi$, 
performed in one of the steps $N(t)$. We have: 
\begin{equation}
 P_\epsilon(x,t) \sim \frac{N(t)}{(x-x_p(t))^{1+\alpha/2}}.
\label{scaling1levy}
\end{equation}
Let us now consider separately the different regimes for the scaling form of the probability distribution.

\begin{figure}
  \includegraphics[width=.7\columnwidth,clip=true]{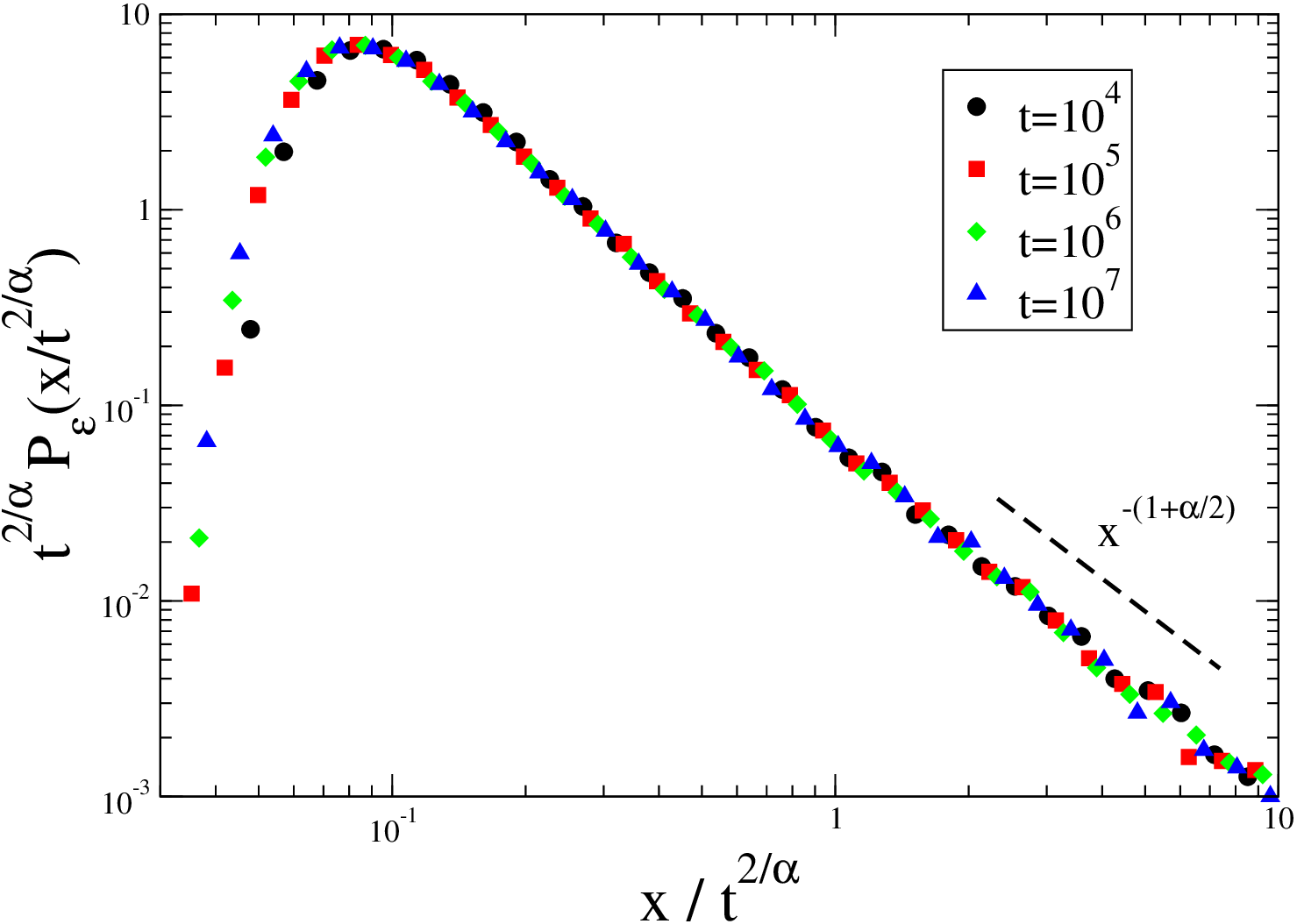}
  \caption{ Probability distribution of displacements $P_\epsilon(x,t)$
    for the Levy walk with exponent {\bf $\alpha = 1/2$} of free
    fights time distribution $p(\tau) = \tau^{-(1+\alpha)}$. The three
    curves (log-log scale) represent data at times $t=10^4,10^5,10^6$,
    collapsed according to the scaling form $P_\epsilon(x,t) = t^{-2}
    F_\epsilon(x/t^2)$.}
  \label{fig:levy-alpha0.5}
\end{figure}

\subsection{Numerical results for the probability distribution: $\alpha < 1$}

In this regime, strong anomalous diffusion is not present and an accelerated growth of 
the scaling length $\ell(t)\sim t^2$ determines the whole dynamics.
In particular also the peak of the distribution grows as 
$x_p(t)\sim t^2$ and
the scaling form in Eq.~(\ref{scaling1levy}) becomes
\begin{equation}
 P_\epsilon(x,t) = \frac{1}{t^2} F\left( \frac{x-x_p(t)}{t^2} \right)=
\frac{1}{t^2} G\left( \frac{x}{t^2} \right),
\label{scaling1alevy}
\end{equation}
This can be clearly seen in Fig.\ref{fig:levy-alpha0.5} at $\alpha =1/2$. 
The scaling form predicted is in excellent agreement with the simulations.
We remark that since the scaling length is of order $t^2$,
it is not possible that a single long jump of order $t^2$ dominates
the transport process.

\subsection{Numerical results for the probability distribution: $1 < \alpha < 2$} 

Numerical simulations show that the peak of the distribution
moves as $x_p(t)= t^{2/\alpha}$
In this case the asymptotic behavior at large $\xi$ 
described in Eq.~(\ref{scaling1levy}) 
is satisfied by the probability distribution and can be used to infer 
its scaling properties: 
\begin{equation}
 P_\epsilon(x,t) = \frac{t}{\xi^{1+\alpha/2} }= \frac{1}{t^{2/\alpha}} F\left( \frac{x-x_p(t)}{t^{2/\alpha}} \right) = \frac{1}{t^{2/\alpha}} G\left( \frac{x}{t^{2/\alpha}} \right),
\label{scaling1blevy}
\end{equation}
with $F(y)=y^{-(1+\alpha/2)}$. Here we used the fact that in this regime
$N(t)\sim t$ and that $x_p(t)$ grows as $t^{2/\alpha}$ in order to
define the new scaling function $G(\cdot)$. 

\begin{figure}[!t]
\includegraphics[width=.7\columnwidth,clip=true]{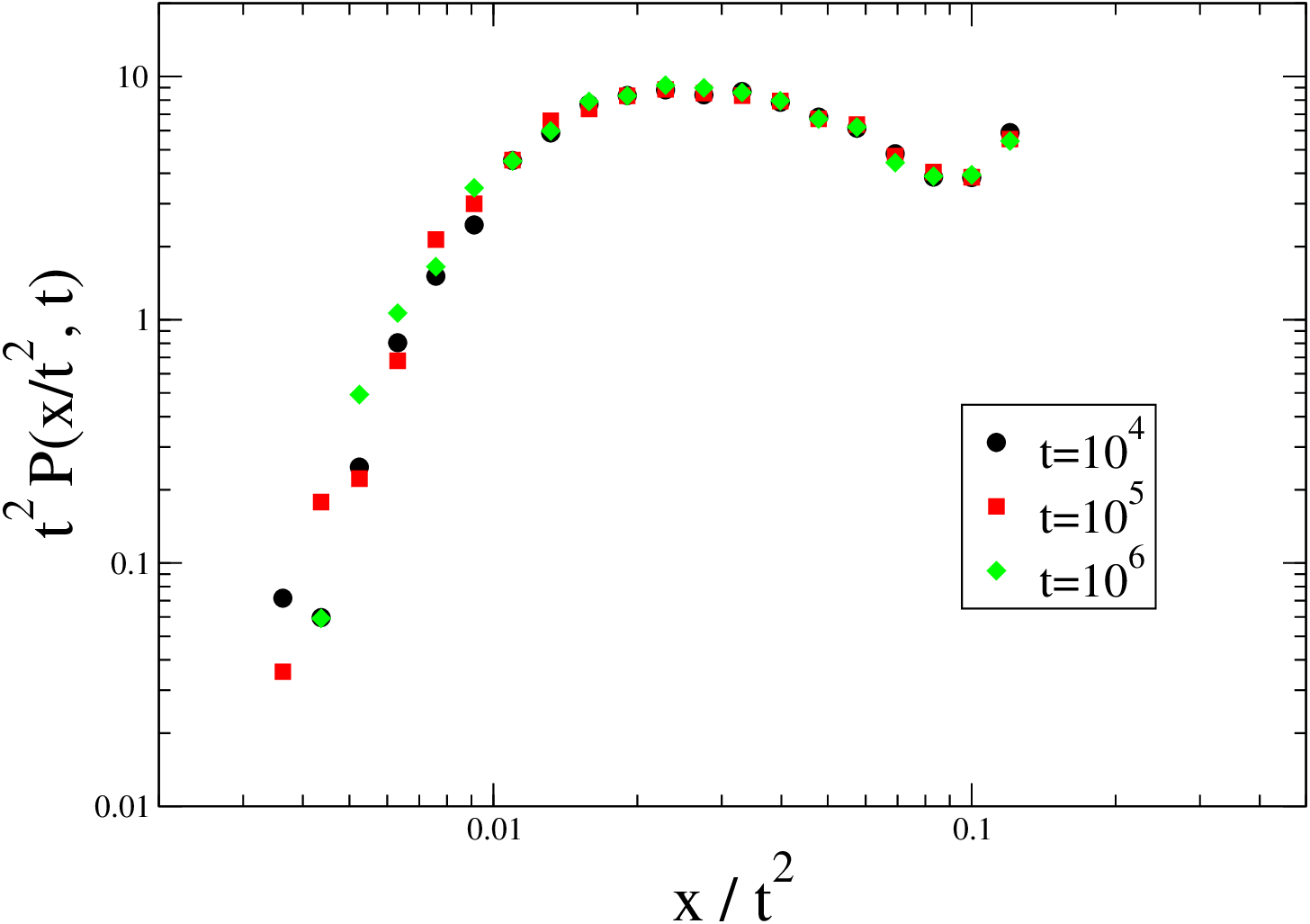}
\caption{ Probability distribution of displacements $P_\epsilon(x,t)$
  for the Levy walk model with bias (log-log
  scale).
  times is obtained with the scaling form in Eq.(\ref{scaling1blevy}),
  with exponent $\alpha=3/2$. Symbols correspond to different times,
  as shown in the legend.  In figure is highlighted the power law
  behaviour $F(y) \sim y^{-(1+\alpha/2)}$ (dashed line) of the scaling
  function $F(y)$ (see text).}
\label{fig:tails-levy-alpha1.5}
\end{figure}

The distribution computed in numerical simulations is shown in
Fig.~\ref{fig:tails-levy-alpha1.5} for $\alpha=1.5$ featuring
a nice data collapse.

\subsection{Numerical results for the probability distribution: $2 < \alpha < 4$}

In this case the peak of the distribution moves linearly $x_p(t)\sim t$
and  the scaling form in Eq.~(\ref{scaling1levy}) becomes
\begin{equation}
 P_\epsilon(\xi,t) = \frac{t}{\xi^{1+\alpha/2} }= \frac{1}{t^{2/\alpha}} F\left( \frac{\xi}{t^{2/\alpha}} \right) = \frac{1}{t^{2/\alpha}} F\left( \frac{x - v t}{t^{2/\alpha}} \right),
\label{scaling2levy}
\end{equation}
with $F(y)=y^{-(1+\alpha/2)}$. 

In the case of the L\'evy walk the superdiffusion is strong anomalous
for all values of the exponent $\alpha$ as we always have
$\langle \left[ \delta x(t)\right]^n \rangle_\epsilon \neq \ell^n(t)$.

\begin{figure}[!t]
\includegraphics[width=.7\columnwidth,clip=true]{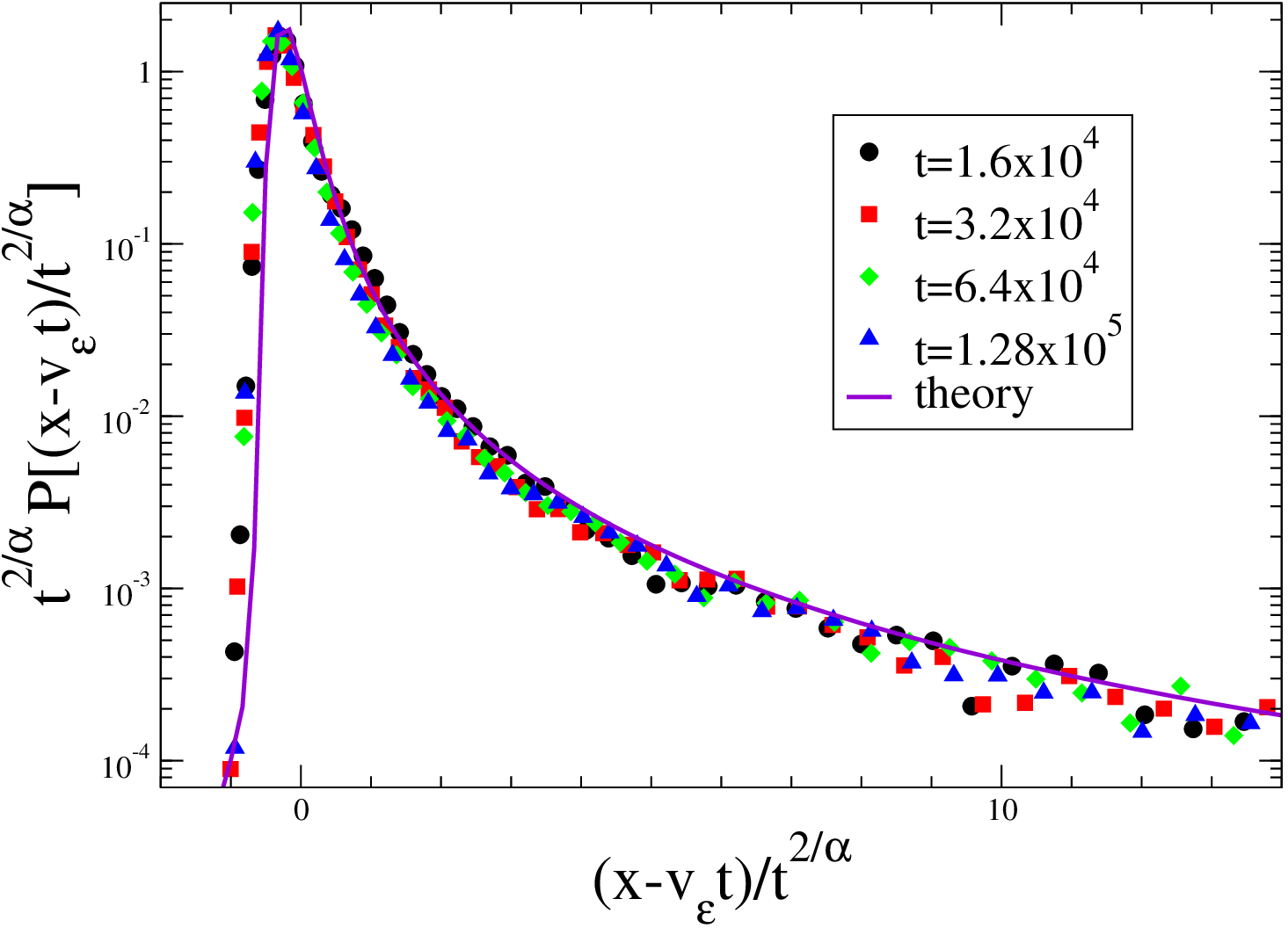}
\caption{
  Probability distribution of displacements $P_\epsilon(x,t)$
  for the Levy walk model with bias (semi-log
  scale).
  The collapse of curves at different times is obtained with the
  scaling form in Eq.(\ref{scaling2levy}), with exponent
  $\alpha=5/2$. Symbols correspond to different times,
  as shown in the legend. The line represents the
  analytical results for the distribution reported in~\cite{noi}.}

\label{fig:tails-levy-alpha2.5}
\end{figure}

Detailed analytical computations are reported in~\cite{noi} and are in
very good agreement with the results of numerical simulations, as
shown in Fig.~\ref{fig:tails-levy-alpha2.5} for $\alpha=2.5$.

\section{Conclusion}

We have studied the scaling properties of the displacement probability distribution functions for a  CTRW with trapping and for the L\'evy walk  in an arbitrary small  field and we have shown that the presence of 
the field introduces new length-scales
related to rare events.  Beyond the principal scaling length of the
distribution $\ell(t)$ and that related to the rigid shift,
$\ell_T(t)$, one must also consider the typical length introduced by
the cut-off of the power-law tail, $\ell_e(t)$, necessary for the
calculation of higher order moments. This is how rare events induce
the \emph{strong} anomalous behavior.  

The scaling properties of probability distribution function and of all its moments can be obtained by estimating the effect of the additional length scales on the process. This has been done by two heuristic yet powerful techniques, the matching condition and the single long jump hypothesis, with excellent agreement with the analytical results obtained from the master equations, and also in excellent agreement with extensive numerical simulations. These heuristic approaches are flexible and could be useful to analyze models where one is not able to solve the master equations for the process in the asymptotic region.

The change in transport properties in the presence of an external field represents an interesting
probe to investigate the microscopic dynamical features of the system.
Field-induced anomalous behavior highlights the importance of rare and
large fluctuations in regimes where the diffusion properties are
\emph{apparently} standard.

\begin{acknowledgments} We thank A. Puglisi for very useful
  discussions. R.B., A.S. and Angelo Vulpiani thank the Kavli Institute for
  Theoretical Physics China at the Chinese Academy of Sciences,
  Beijing, for the kind hospitality at the workshop "Small system
  nonequilibrium fluctuations, dynamics and stochastics, and anomalous
  behavior" July 2013 where the paper was completed. The work of G.G.
  and A.S. was supported by the Granular Chaos project, funded by the
  Italian MIUR under the grant number RIBD08Z9JE.
\end{acknowledgments}

\end{document}